\documentclass[twocolumn,showpacs,preprintnumbers,amsmath,amssymb,aps,prl,superscriptaddress]{revtex4-2}
\usepackage[title]{appendix}
\usepackage{graphicx}
\usepackage[english]{babel}
\usepackage{booktabs}
\usepackage{dcolumn}
\usepackage{bm}
\usepackage{url}
\usepackage{natbib,hyperref}
\usepackage{amsmath}
\usepackage{xfrac}
\usepackage[T1]{fontenc}
\usepackage[usenames,dvipsnames]{xcolor}

\setcitestyle{numbers, square}
\DeclareMathOperator*{\argmax}{arg\,max}\DeclareMathOperator*{\argmin}{arg\,min}

\newcommand{\EB}[1]{\textcolor{black}{#1}}
\newcommand{\CH}[1]{\textcolor{black}{#1}}
\newcommand{\AP}[1]{\textcolor{black}{#1}}

\begin{document}
\title{Learning complexity to guide light-induced self-organized nanopatterns}

\author{Eduardo Brandao}
	\affiliation{Univ Lyon, UJM-Saint-Etienne, CNRS, IOGS, Laboratoire Hubert Curien UMR5516, F-42023 St-Etienne, France}
\author{Anthony Nakhoul}
	\affiliation{Univ Lyon, UJM-Saint-Etienne, CNRS, IOGS, Laboratoire Hubert Curien UMR5516, F-42023 St-Etienne, France}
	\author{Stefan Duffner}
\affiliation{Univ Lyon, CNRS, INSA-Lyon, LIRIS, UMR5205, France}
\author{Rémi Emonet}
	\affiliation{Univ Lyon, UJM-Saint-Etienne, CNRS, IOGS, Laboratoire Hubert Curien UMR5516, F-42023 St-Etienne, France}
	\author{Florence Garrelie}
\affiliation{Univ Lyon, UJM-Saint-Etienne, CNRS, IOGS, Laboratoire Hubert Curien UMR5516, F-42023 St-Etienne, France}
\author{Amaury Habrard}
\affiliation{Univ Lyon, UJM-Saint-Etienne, CNRS, IOGS, Laboratoire Hubert Curien UMR5516, F-42023 St-Etienne, France}
\affiliation{Institut Universitaire de France (IUF), Paris, France}
\author{François Jacquenet}
\affiliation{Univ Lyon, UJM-Saint-Etienne, CNRS, IOGS, Laboratoire Hubert Curien UMR5516, F-42023 St-Etienne, France}
\author{Florent Pigeon}
\affiliation{Univ Lyon, UJM-Saint-Etienne, CNRS, IOGS, Laboratoire Hubert Curien UMR5516, F-42023 St-Etienne, France}
\author{Marc Sebban}
\affiliation{Univ Lyon, UJM-Saint-Etienne, CNRS, IOGS, Laboratoire Hubert Curien UMR5516, F-42023 St-Etienne, France}
\author{Jean-Philippe Colombier}
\email{jean.philippe.colombier@univ-st-etienne.fr}
\affiliation{Univ Lyon, UJM-Saint-Etienne, CNRS, IOGS, Laboratoire Hubert Curien UMR5516, F-42023 St-Etienne, France}
    
\date{\today}

\begin{abstract}
\EB{Ultrafast laser irradiation can induce spontaneous self-organization of surfaces into dissipative structures with nanoscale reliefs. These surface patterns emerge from symmetry-breaking dynamical processes that occur in Rayleigh-Bénard-like instabilities. In this study, we demonstrate that the coexistence and competition between surface patterns of different symmetries in two dimensions can be numerically unraveled using the stochastic generalized Swift-Hohenberg model. We originally propose a deep convolutional network to identify and learn the dominant modes that stabilize for a given bifurcation and quadratic model coefficients. The model is scale-invariant and has been calibrated on microscopy measurements using a physics-guided machine learning strategy. Our approach enables the identification of experimental irradiation conditions for a desired self-organization pattern. It can be applied generally to predict structure formation in situations where the underlying physics can be approximately described by a self-organization process and data is sparse and non-time series. Our work paves the way for supervised local manipulation of matter using timely-controlled optical fields in laser manufacturing.}
\end{abstract}

\maketitle
\EB{The emergence of instabilities and symmetry breaking leading to the formation of coherent structures, is one of the most fascinating aspects of the complex dynamics governing light-surface interaction~\cite{her1998microstructuring,shimotsuma2003self,ilday2017rich}.
When a randomly rough surface is subjected to ultrafast laser pulses, it enters a far-from-equilibrium state due to the repeated absorption of pulsed optical fields. As a result, the surface exhibits spontaneous spatial organization, which is oriented by energy gradients generated by laser polarization, giving rise to laser-induced periodic surface structures (LIPSS)~\cite{sipe1983laser}. These structures form under far-from-equilibrium conditions and can be triggered by capillary waves, convection rolls, and thermoconvective instabilities,~\cite{keilmann1983laser,young1984laser,tsibidis2016convection,rudenko2020high} which persist through dissipative structures~\cite{prigogine1963introduction}.  Eliminating the prevailing laser polarization effects reveals puzzling patterns emerging from a sequence of instabilities, inducing different types of complex patterns, ranging from chaos to six-fold symmetries~\cite{Nakhoul2021Apr}. The photoexcited matter undergoes a transition from a disordered state to a more coherent one, referred to as a \emph{strange attractor} in the phase space of nonlinear dynamics. This transition results in a metastable state, defining a self-organization structuring regime. Through this self-organization process, the material surface can be sculpted seamlessly, enabling nanoscale manufacturing~\cite{nakhoul2022boosted}. Understanding the selection mechanisms involved in this morphogenesis to gain control over the uniformity, symmetry, and size of the resulting surface patterns is a major research theme in laser processing for photonics metasurfaces, biomimetics, or catalysis functionalization.~\cite{stratakis2020laser,overvig2020multifunctional}. To apply statistical inference approaches to complex systems and achieve generalizability, advanced physics-guided machine learning strategies are essential.}

Upon laser irradiation, a hazy boundary separates self-organized and organized surface patterns. When a material is exposed to sufficiently intense laser irradiation, it tends to organize along the stationary electromagnetic fields due to scattered/excited waves~\cite{sipe1983laser,rudenko2019self} and self-organize in response to the random fluctuations of light absorption with a symmetry breaking with respect to polarization~\cite{varlamova2006self,abou2019nanoscale}. Light-oriented and self-assembled dynamical processes are inherently superimposed, and surface topographies evolve spatio-temporally towards equilibrium patterns that result from a complex competition between free energy dissipation imposing entropy production and spontaneous ordering.

Consequently, any preexisting or transient organization can be disrupted by random perturbations, which can be amplified by positive feedback to lead the system towards new patterns.

Ultrafast laser texturing has recently been used to obtain deep sub-wavelength periodic patterns, which raises questions about the relevant electromagnetic processes that drive the formation of these patterns well below the diffraction limit~\cite{stoian2020advances,bonse2020maxwell}.

\EB{Various types of 2D surface patterning have been reported, including patterns with oriented, triangular, hexagonal, labyrinthine, or chaotic symmetries~\cite{qiao2018formation,fraggelakis2019controlling,AbouSaleh2020Mar,mastellone2021deep}, featuring both positive and negative reliefs such as humps, bumps, peaks, and spikes~\cite{Nakhoul2021Apr}. To explain the remarkably uniform establishment of these patterns on the microscale independently from the oriented near-field optical effects on the random local nanotopography, a more global and collective perspective is required~\cite{AbouSaleh2020Mar,Nakhoul2021Apr}. Nanoscale fluid flows were shown to be driven by a complex interplay between electromagnetic, internal and surface pressure forces which can become trapped due to the resolidification process~\cite{AbouSaleh2020Mar,rudenko2020high}}

\EB{The deterministic approach to predict the underlying optical coupling processes is limited because it requires the artificial integration of fluctuating conditions induced by surface roughness. Transiently formed structures can become unstable under nonlinear amplification and bifurcate into more complex patterns that are not accurately described by classical approaches like Navier-Stokes combined with Maxwell equations. Nonetheless, the complex pattern landscape has been experimentally explored and can be now compared with mathematical models dedicated to nonlinear system dynamics.
}

\EB{The Kuramoto-Sivashinsky approach has become a paradigm for describing pattern formation and spatiotemporal chaos on surfaces eroded by ion bombardment, which ultimately reproduces ripple formation and other organized patterns~\cite{bradley2010spontaneous}. A similar approach was initially proposed for laser-induced nanopatterns, although a clear physical picture has yet to be established~\cite{reif2012role}. Along similar lines, the Swift-Hohenberg (SH) dynamics has been identified as a relevant candidate for representing the observed complexity of convective instabilities with spatiotemporal features, such as chaos, rolls, and hexagons~\cite{elder1992ordering,cross1993pattern}. The SH approach has proven to be useful in identifying generic spatiotemporal dynamics of patterns in convective fluids~\cite{decker1994spiral,echebarria2000defect}, as well as curvature- and stress-induced pattern-formation transition~\cite{stoop2015curvature}. The SH approach was formally deduced from the Navier-Stokes equations in the Boussinesq approximation, with thermal fluctuation effects in a fluid near the Rayleigh-Bénard instability~\cite{swift1977hydrodynamic}.}

\EB{The purpose of this letter is to demonstrate that laser-induced pattern formation at the nanoscale can be efficiently characterized and predicted by a stochastic SH model that is variational in time and conservative in space.
Our original strategy relies on the use of machine learning (ML) integrating partial physical information in the form of the SH model, which allows us to identify dominating stable modes for a set of parameters \emph{independently} of initial roughness conditions. 
Incorporating data and prior knowledge is naturally expressed in terms of Bayesian inference, for which well-established domain-specific methods exist dating back to Laplace~\cite{tarantola05_inver}, but which cannot be applied in our experimental situation of few data and partial physical knowledge: in geophysics and climate science, where the physical process is well-understood, methods focus on state reconstruction, known as \emph{data assimilation}~\cite{carrassi18_data_assim_geosc}; in physics, since states can be prepared, \emph{model calibration} was developed~\cite{kennedy01_bayes_calib_comput_model}, with recent advancements using ML~\cite{viana21_survey_bayes_calib_physic_infor} to integrate the parameters of either the full model or a correction to incomplete physical knowledge from data~\cite{yin21_augmen_physic_model_with_deep}. However, solving the \emph{joint} inverse problem of finding both state and model parameters is more challenging. In the climate sciences, sophisticated machine-learning techniques were recently proposed, integrating physical information via constraints, either during training or in model architecture itself~\cite{beucler19_achiev_conser_energ_neural_networ, filoche20_compl,farchi21_using_machin_learn_to_correc,nguyen20_assim,dechelle20_bridg_dynam_model_deep_networ}, but require abundant time-series data. 
Our original strategy allows us to solve the dual inverse problem using only \textit{one} observed state --- a scanning electron microscope (SEM) image --- even with little data. Furthermore, our modelling is scale-invariant and can be applied to \textit{any} laser process. By reducing experimental irradiation parameters to simple model coefficients, they can be optimized and extrapolated for surface pattern engineering.}
\begin{figure}[h!]
\centering\includegraphics[width=8.5 cm]{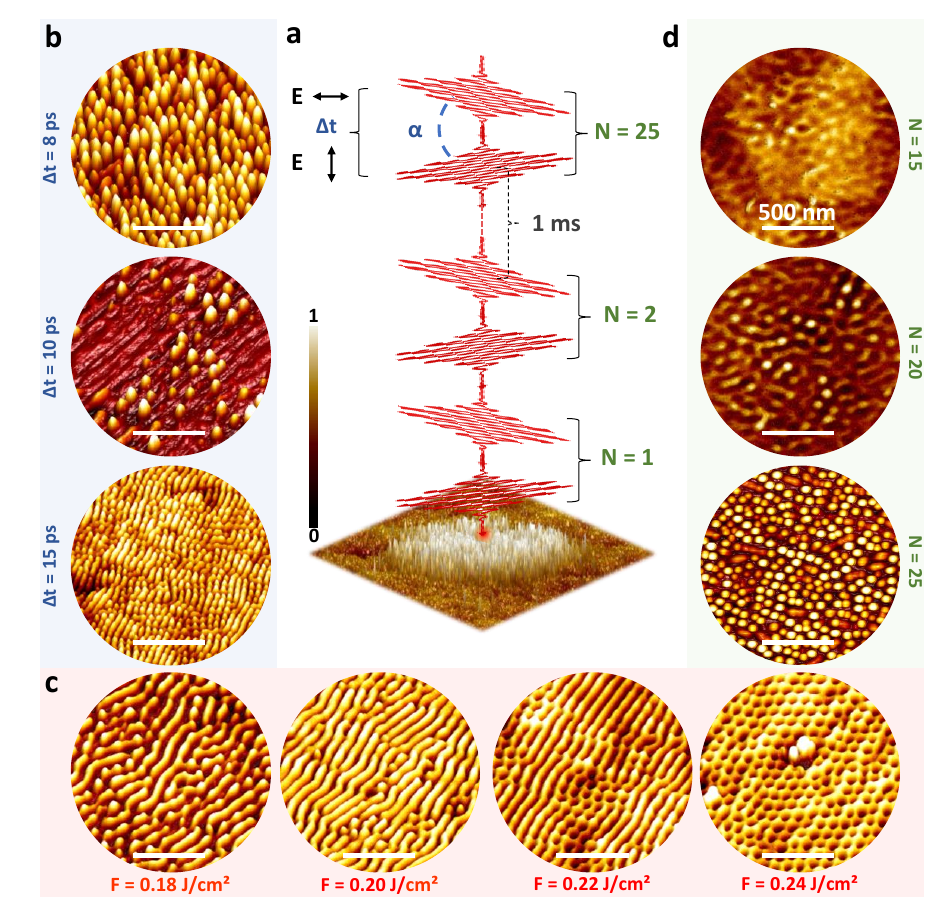}
\caption{\label{fgr1} (a) Schematic illustration of experimental self-organization regimes induced by bursts of ultrafast laser (150 fs) double pulses. (b) Self-organized patterns of topography that develop varying time delays for a given $F$ and $N$ (AFM-3D mode). (c) Nanopattern variation with respect to laser fluence at fixed $\Delta t$ and $N$ (AFM-3D mode). (d) Nanostructure growth by feedback at different number of pulses (AFM-2D mode), for a fixed $\Delta t$ and $F$. The scale bars represent a length of 500 nm.}
\end{figure}

Tailoring nanotopographic features on a surface is a challenging task that has been successfully accomplished using ultrafast laser processes with time-controlled polarization strategies. Numerous regimes of LIPSS have been reported with various periodicities, heights, orientations, and symmetries depending on different polarization directions between the first $\vec{E_{1}}$ and second pulse $\vec{E_{2}}$, characterized by $\alpha=$($\vec{E_{1}}\cdot\vec{E_{2}}$) in Fig.\ref{fgr1}(a)~\cite{Bonse2012Jul,Wang2020Apr,AbouSaleh2020Mar,Nakhoul2021Apr}.

\EB{Figs.\ref{fgr1}(b-d) present surface topographies measured by high resolution atomic force microscopy (AFM). A circular region with a diameter of 1 $\mu$m corresponding to the laser impact center was mapped in 3D (tilted) mode in Fig.\ref{fgr1}(b-c) and in 2D for Fig.\ref{fgr1}(d). To observe the significant role of temporal pulse splitting $\Delta t$ in nanopatterns control, laser peak fluence $F$ and $N$ were kept fixed at 0.18 J/cm$^{2}$ and 25 respectively, as shown in Fig.\ref{fgr1}(b). At $\Delta t$ = 8 ps, organized nanopeak structures were observed with a high aspect ratio, a height of $\sim$ 100 nm and a diameter of $\sim$ 20 nm~\cite{nakhoul2022boosted}. An extension of 2 ps in $\Delta t$ modifies the observed patterns that turn into a different organization, a regime referred to as nanobumps~\cite{Nakhoul2021Apr}. For $\Delta t$ = 15 ps, a regime of nanohump generation is reached with a lower aspect ratio as the structures display a height of $\approx$ 10 nm and a diameter of $\approx$ 30 nm. }

\EB{The role of laser fluence is revealed by fixing $\Delta t= 25$ ps and $N= 25$, as depicted in Figure \ref{fgr1}(c). At $F=0.18$ J/cm$^{2}$, a low-contrast nanopeak regime is formed, evolving into a nanostripe pattern with a slight increase in laser fluence increase to 0.20 J/cm$^{2}$. At $F=0.22$ J/cm$^{2}$, a transition region is established, combining both stripes and cavities. Finally, at $F=0.24$ J/cm$^{2}$, the surface is uniformly organized with hexagonally arranged nanocavities having a depth of $\approx$ 25 nm and a diameter of $\approx$ 30 nm. Both nanohumps and nanovoids result from hydrothermal flows guided by surface tension and rarefaction forces, leading to thermoconvective instability at the nanoscale, similarly to well-known Rayleigh-Bénard-Marangoni instabilities~\cite{AbouSaleh2020Mar,Nakhoul2021Apr,Vitral2020Sep,Bodenschatz2000Jan,Thess1995Dec,Pearson1958Sep,Morgan2018Mar,Smith1983Jul,Smith1986Oct,Tsibidis2015Jul,Busse2014Sep,Boeck1999Nov,Bragard1998Aug,Starikov2015Apr,rudenko2020high}. Laser dose also plays a role, as positive feedback regulates pulse-to-pulse topographical transformations. As shown in Fig.\ref{fgr1}(d), at a fixed $F=0.24$ J/cm$^{2}$ and  $\Delta t =8$ ps with different $N$, corresponding to the parameters of nanopeaks formation presented in Fig.\ref{fgr1}(a), three different surface organizations were observed. Pulse-to-pulse growth dynamics exhibits the transitions from convection cells ($N$ = 15), to the creation of crests on the convection cells ($N$ = 20). The nanopeaks grow on the edges of the crests to reach their optimal shape, concentration and organization at  $N$ = 20.}

The adimensional form of the generalized Swift-Hohenberg equation (SH) used in this letter is (see derivation in Suppl. Mat.): \begin{equation}
\dot{\tilde{u}} = \epsilon \tilde{u} - (1+ \tilde{\nabla}^2)^2\tilde{u} + \gamma \tilde{u}^2 - \tilde{u}^3.
\end{equation}
The SH model was introduced in~\cite{swift1977hydrodynamic} as a model of Rayleigh-Bénard convection, modified by the inclusion of a $u^2$ nonlinearity allowing for small amplitude destabilization and the emergence of experimentally observed hexagonal patterns.

With appropriate boundary conditions, the original SH equation exhibits a type-I-s instability that is isotropic, invariant with respect to translations and to \(u\rightarrow -u\)~\cite{cross1993pattern}. Perturbations of \(u_{b}=0\) are selectively amplified depending on the norm of the wave number, leading to the formation of complex \emph{patterns} with no preferential direction.
\EB{The generalized} SH model has the Lyapunov functional \(\mathcal{L} \left[ \tilde{u}  \right] = \int_{\Omega} \frac{\tilde{u} }{2}\left( \nabla^4\tilde{u}  + 2\nabla^2\tilde{u} +\tilde{u}  \right)+\frac{1}{4}\tilde{u} ^4 -\frac{\gamma}{3} \tilde{u} ^3-\frac{\epsilon}{2}\tilde{u} ^2 d\mathbf{x}\) and  \(\dot{\tilde{u} } = - \frac{\delta \mathcal{L}}{\delta \tilde{u} }\), as can be readily verified.

During the SH dynamics, the Lyapunov functional $\mathcal{L}$ decreases in the same way as entropy decreases during the formation of physical patterns, and it converges asymptotically to a stable value~\cite{cross1993pattern} (see Fig.~\ref{fgr2}).
\begin{figure}[h!]
\includegraphics[width=8.5 cm]{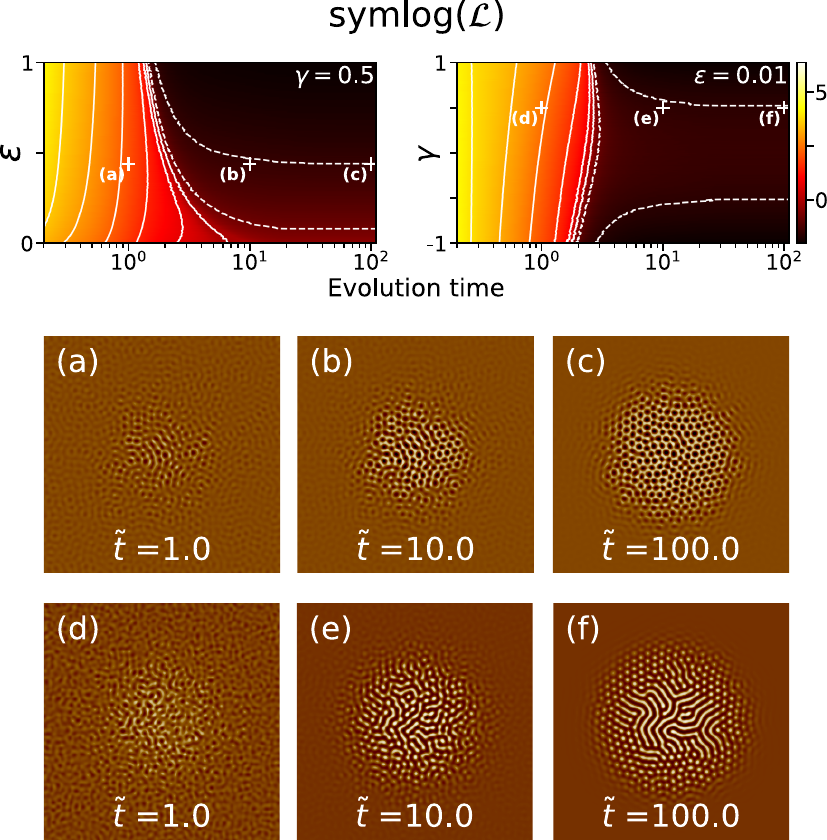}
\caption{\label{fgr2}Lyapunov functional of the generated field solutions of the SH equation as a function of evolution time $\tilde{t}$ for fixed $\epsilon$ and $\gamma$ ($\epsilon$ a centered 2D Gaussian ramp to mimic the laser fluence distribution), depicted as a heatmap in symlog scale, for independent initial conditions. Lyapunov functional evolution is largely independent of initial conditions and decreases during dynamics. The SH equation is able to reproduce, among others, highly symmetric hexagonal solutions (top), as well as labyrinthine solutions surrounded by nanopeaks.
}
\end{figure}

We numerically solve the SH equation using a second-order Strang splitting pseudo-spectral solver with an adaptive time step~\cite{strang1968construction,leveque2007finite,yoshida1990construction,cooley1965algorithm,butcher2016numerical}, offering a good compromise between accuracy and speed.
Fig.~\ref{fgr2} a-c and Fig.~\ref{fgr2} d-f show evolution dynamics of pattern formation for two pairs of $\epsilon, \gamma$.

\begin{figure}[h!]
\centering\includegraphics[width=8.6 cm]{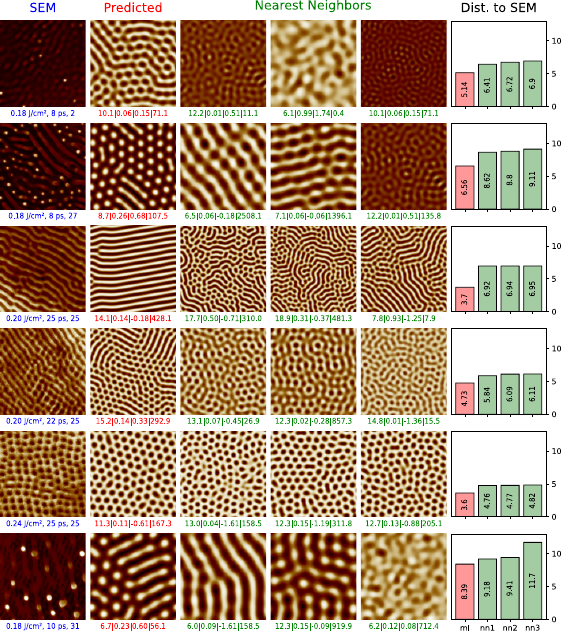}
\caption{\label{fgr3} \EB{Each row shows a 224 by 224 pixel SEM experimental image, with $1 \mu m\approx237$ pixels which was never seen by the ML model during learning, the corresponding ML-predicted image for the same laser parameters, and three nearest neighbors (NN) of the former among solver generated images; and a bar plot of $l_2$ distance in the image of the feature mapping $F_m$ between SEM image, ML-predicted image (ml) and NN (nn1, nn2, nn3). Image labels, left to right: $F, \Delta t, N$ (SEM); predicted SH parameters $l_p$, $\epsilon_p$, $\gamma_p$, $\tilde{t}_p$ (other). Bar plots: ML predictions are more accurate, as distance between SEM and ML predictions is smaller than to NN, the former integrating global information. On the first and second rows, NN with different length scales can be observed, suggesting concurrent multi-scale SH processes. The ML model, which integrates single-scale SH knowledge, can only predict one of these processes.}}
\end{figure}

A ML model is employed to learn the relationship between observed laser parameters \(\theta\) and patterns, using only few, non-time series data (I), assuming an approximately SH process, not explicitly given in terms of $\theta$, parameterized by $\varphi$ (consisting of a scale factor $l$, the maximum wavenumber in a domain of side 224 pixels given as multiple of $2\pi$, the adimensional model parameters $\epsilon$ and $\gamma$, and $\tilde{t}$, which can be seen as a stabilization time) (II), with unknown initial conditions $u_0$\ (III). 

\EB{We motivate this choice by symmetry considerations (Suppl. Mat.) as well as the similarity between SEM images and SH solutions (Fig.\ref{fgr3}). Combining experimental information with that obtained via the ML model, we find that the timescale of the convective instability is consistent with that reported in~\cite{Nakhoul2021Apr} (Suppl. Mat.), further supporting our choice}.
Learning the relationship between laser parameters and patterns consists in solving the dual inverse problem of estimating an unknown initial state and model parameters with severe constraints, which is a challenging task and cannot be tackled in general using only ML methods. However, for a self-organization process, stating initial conditions exhaustively is wasteful, since for random perturbations of the uniformly zero solution of SH most Fourier modes are attenuated. A \emph{feature mapping} \(F_m\) is therefore defined that is simplifying (non injective) and discriminating (if \(u^{i}, u^{j}\) have different patterns then \(F_m(u^{i})\neq F_m(u^{j})\)) such that the image of the data distribution under $F_m$ is conditionally independent of $u_0$ given the physical knowledge $\varphi$. \EB{This considerably simplifies the problem since the initial state no longer needs to be estimated.} Learning $F_m$~\cite{greydanus19_hamil_neural_networ} from few data is impractical~\cite{vapnik99}, (II) precludes deriving ite on first principles, and using traditional image features would limit discriminating power for unknown patterns. $F_m$ is therefore chosen as a deep convolutional neural network~\cite{lecun1998gradient} (CNN) pretrained for a broad classification task on Imagenet~\cite{deng09_imagen}, since CNNs are translation equivariant (making them suited for a pattern specification task). Their features are learned automatically from data, and retain scale information~\cite{graziani21_scale_invar_state_art_cnns_train_imagen}. Given experimental data \({\left\{ \theta^i,u^i \right\}}_{i=1\ldots N }\), we learn \(\tilde{\varphi}_\alpha\) that maximizes the log likelihood of the observed \(F_m(u^{i})\):
\begin{align}
\label{eq:learn-via-max-log-likelihood-obs}
\bar{\alpha}=\argmax_{\alpha} \sum\limits_{i=1}^{N}\log p \left( F_m(u^{i})|\tilde{\varphi}_\alpha(\theta^{i}) \right)
\end{align}
Assuming that the distribution of $\varphi$ given $\theta$ is peaky, we label experimental $u^i$ with $\tilde{\varphi}^i$ the SH parameters of its nearest neighbor (NN), in the image of $F_m$, among a large number of $u$ pre-generated with the SH solver from random $u_0$. By integrating physical knowledge in this way, the problem of maximizing the likelihood above can be replaced with a lower bound. Explicitly, assuming data is generated i.i.d. from a Gaussian distribution,
\begin{align}
    \label{eq:expectation-maximization}
    \bar{\alpha}=\argmin_{\alpha}\frac{1}{N}\sum\limits_{i=1}^N \left\|\tilde{\varphi}^i-\tilde{\varphi}_{\alpha}(\theta^{i})  \right\|^{2}
\end{align}
which is a low-dimensional problem that can be solved with few data~\cite{bishop2006pattern} with a support vector regressor~\cite{drucker1996support} $\tilde{\varphi}_\alpha$ parameterized by $\alpha$.
dimensional problem, which can always be solved with fewer data.
\begin{figure}[h!]
\includegraphics[width=8.6 cm]{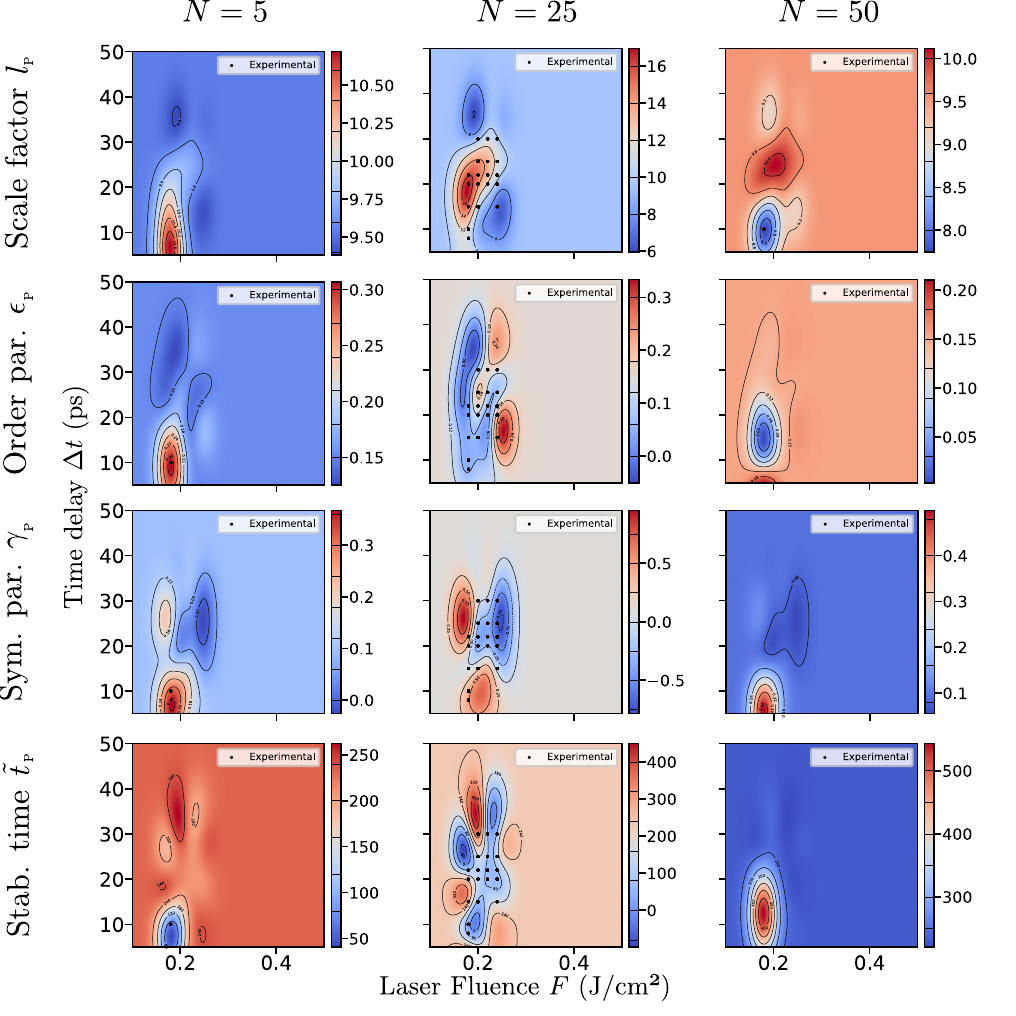}
\caption{\label{fgr4} Each plot shows, \EB{as a heatmap,} the ML model prediction of a single SH parameter (bottom to top: scale factor $l_p$; order parameter \(\epsilon_p\); symmetry breaking parameter \(\gamma_p\); simulation stabilization time $\tilde{t}_p$) as a function of laser fluence, time delay, and number of pulses (respectively, x-axis and y-axis, and column). Experimental points are overlaid on each plot.}
\end{figure}

\CH{Figure~\ref{fgr3} demonstrates the remarkable accuracy of our ML strategy in predicting the shape and scale of experimental patterns, even for never-before-seen laser parameters. Our strategy is more efficient than local methods that rely on nearest neighbor information, since the distance to the SEM experimental patterns in the image of the feature mapping $F_m$ is smaller. As shown in Fig.~\ref{fgr4} the complexity of the learned relationship between laser parameters and SH parameters grows with the number of experimental observations, with sharp boundaries of rapidly varying parameter values in regions of many data. The ML model can extrapolate to regions with few data for $N$ and $\Delta t$, but less so for $F$, which would require higher experimental resolution.} Importantly, we find that predicted SH parameters are correlated, and the correlation sign changes with $N$ (Suppl. Mat.): $l_p$, for example, is inversely correlated with pattern characteristic size and increases with $N$ (the increase is not uniform, being greater for large $l_p$ regions).

\EB{This parameter is particularly important as the characteristic size of a stable mode is of great interest for applications. Because $l_p$ and other parameters are correlated, it cannot be set freely; but as seen in Fig.~\ref{fgr4} parameter isosurfaces are orthogonal at places: at e.g. $N=25$, a high-gradient transition regime for $l_p$, at $F=0.18$ J/cm$^{2}$, $\Delta t = 15$ ps in the $\Delta t$ direction is observed, while for $F, \Delta t$ in the same region, the other SH parameters remain roughly constant. Varying $l_p$ in the direction of $\nabla l_p$ thus allows adjusting the characteristic size of the particular stable mode defined by the other SH parameters. This opens the door to pattern optimization for specific applications.}

Interestingly, ($\gamma_p$) determines whether holes or bumps are observed; $F=0.2$ J/cm$^{2}$ separates a region of high and low $\gamma_p$, roughly independently of $N$; the sign of $\gamma_p$ appears to be determined by $F$ and $\Delta t$ only. Furthermore, for large bifurcation parameter ($\epsilon_p$), many modes are non-attenuated and patterns are less ordered. Correspondingly, large $\epsilon_p$ patches are observed at high $F$/low $\Delta t$ (highest energy coupling). For $N=25$, superimposing the $\gamma_p=0$ isosurface on the $\epsilon_p$ prediction, it can be seen that it is roughly perpendicular to isosurfaces of $\epsilon_p$. These abrupt transition regimes of $\epsilon_p$ are consistent with experimental observations where \textit{two} patterns of different order are superimposed on the \textit{same} SEM image. ($\tilde{t}_p$) for constant $l_p, \epsilon_p, \gamma_p$, symmetry increases with $\tilde{t}_p$, as symmetrical patterns require large $\tilde{t}_p$ to stabilize from a uniformly random state. As can be seen in the bottom row, $\tilde{t}_p$ tends to increase with $N$, consistently with the physical view that a large $N$ increases the time the dissipative system is in a far from equilibrium state. This increase is not uniform across $F, \Delta t$ pairs, and the area of laser parameter space of relatively large $\tilde{t}_p$ decreases with $N$.

We show that ultrafast laser-irradiated surface nanoscale patterns can be numerically modelled by a scale-invariant generalized Swift-Hohenberg equation. A machine learning model is trained to learn the connection between the stochastic SH equation and laser parameters, independently of initial conditions, using a deep convolutional network to extract features and by incorporating physical information. 
\EB{Our original strategy can generally be applied to accurately predict the shape and scale of physical patterns generated by other self-organization processes, even if the underlying physical model is only approximate and experimental data is limited and non-time series.}
The ML model is able to identify regions of laser parameters that are relevant for applications and can even be used to predict novel patterns, since the convolutional neural network features are not learned from observed patterns. Regions where pattern superpositions are observed could be modeled more accurately via a mixing of SH processes, as a manifestation of superposed states of self-organization, providing new routes toward nanoscale surface manipulation by light.

This work has been funded by a public grant from the French National Research Agency (ANR) under the “France 2030” investment plan, which has the reference EUR MANUTECH SLEIGHT -ANR-17-EURE-0026.

\AP{\paragraph{Appendix on experimental set-up.} In the proposed experiment, Mach-Zehnder interferometry was used to combine the effect of polarization mismatch with an adjustable inter-pulse delay $\Delta t$, enabling fine control of surface topography at the tens of nanometer scale~\cite{Nakhoul2021Apr}. By breaking the surface isotropy imposed by a single polarization state, a wide range of self-organization regimes was achieved on a nickel monocrystal oriented in the (001) direction. Specifically, using a cross-polarization strategy, setting a depolarization angle of $\alpha =90^\circ$ and a range of time delays between 8 and 25 ps were set, as shown in Fig.\ref{fgr1}(a). The pulse duration was fixed at 150 fs, and the laser dose was finely controlled by the number $N$ of double-pulse sequences. Prior to laser irradiation, the Nickel surface was mechanically polished with a Ra $<5$ nm to ensure that the surface dynamics followed a hydrodynamics-governed process, smoothing the inhomogeneous electromagnetic response.}
\AP{\paragraph{Appendix on Image similarity.} It is important to note, regarding the ML strategy, that the problem is approached within the image of a feature mapping $F_m$, where the concept of similarity differs from \textit{visual} similarity. In this feature space, an image is equivalent to any of its images in the orbit of the group of symmetries of $F_m$, such as translations, but also a variety of other symmetries that are learned from data automatically. Intuitively, similarity in feature space corresponds to similarity of \textit{patterns}, which can be described in terms of e.g. "bumpiness," "roundness," etc.}

\bibliography{SHEbib}

\begin{appendices}
\section{Appendix A: Experimental self-organization regimes}
\paragraph{Delay, Fluence and number of double-pulses dependence}
\begin{figure}[h!]
\includegraphics[width=8.15cm]{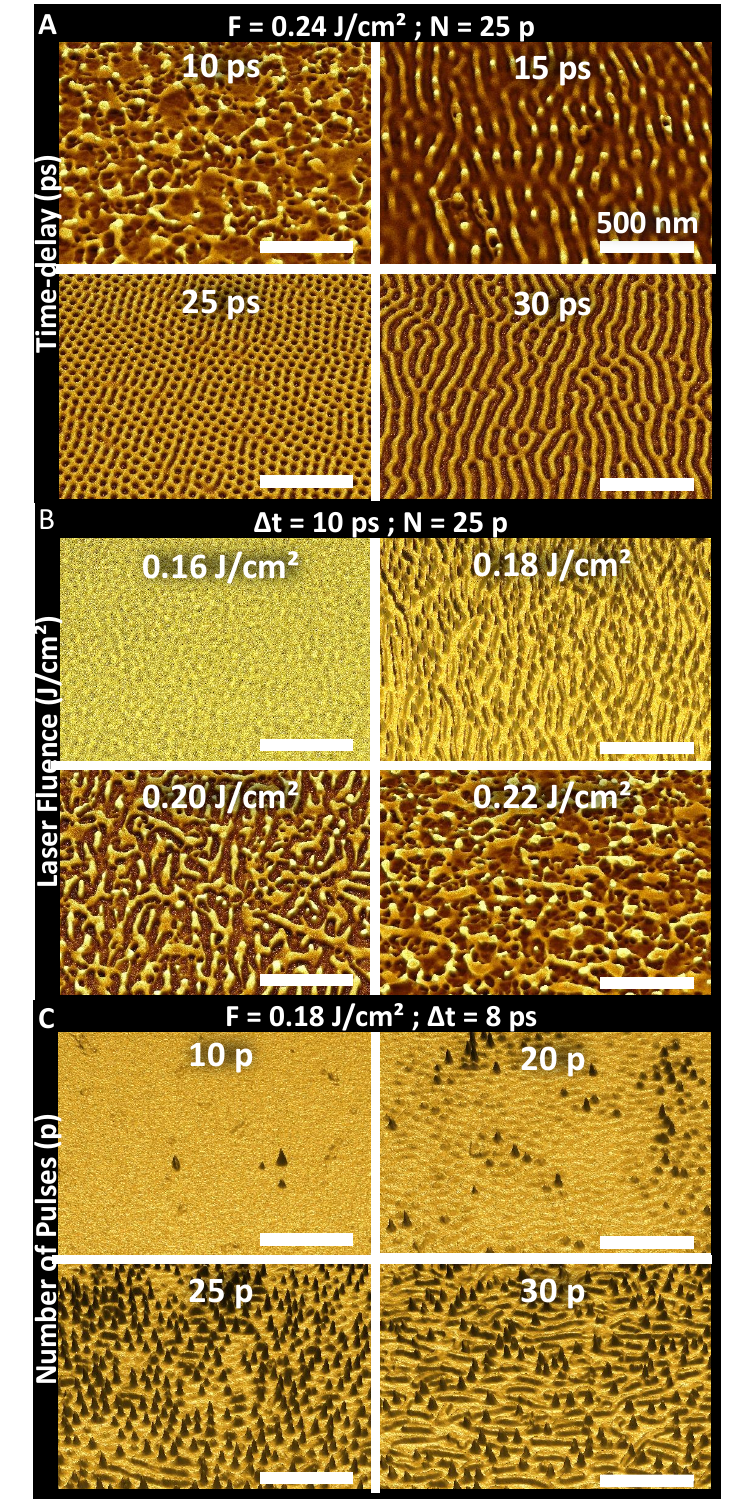}
\caption{\label{fgrexp} SEM images of self-organization regimes obtained by ultrafast laser irradiation of nickel samples. The final patterns as chaos, hexagonal arrays, oriented stripes, labyrinthine structures and peaks are determined by the time-delay ($\Delta t$) between the two cross-polarized pulses (A), the laser fluence ($F$) (B) and the number of pulses ($N_{DPS}$) (C).}
\end{figure}
Scanning electron microscopy images (SEM) of several organization regimes are presented in Fig.\ref{fgrexp}(A) controlled by different time delay $\Delta t$ between the two cross-polarized pulses at a fixed number of double-pulses sequences  $N_{DPS}=25$ and fixed laser fluence of $F=0.24$ J/cm$^{2}$. We observe that time delay finely controls the surface topography. By progressive increases of $\Delta t$ by 5 ps time lapses, the nanostructures organizations and morphologies are tuned. For example, self-organized nanocavities are observed at $\Delta t=25$ ps. For larger time-delay of 30 ps, the hexagonal nanocavities are transformed into labyrinthine nanostructures. Thus, the time delay is one of the key parameters to regulate the photon flux driving the formation of selected self-organized nanopatterns.

Fig.\ref{fgrexp}(B) presents the formation of analogous surface topography by only varying the laser fluence between $F=0.16$ J/cm$^{2}$ and $F=0.22$ J/cm$^{2}$, at a fixed $N_{DPS}=25$ ps and $\Delta t=10$ ps. At $F=0.16$ J/cm$^{2}$, some nanoroughness started to appear on the irradiated surface. For $F=0.18$ J/cm$^{2}$, a forest of organized nanopeaks was observed on the irradiated surface. At higher laser fluence between 0.20 and 0.22 J/cm$^{2}$, the nanopeaks structures turned to chaotic structures. Therefore, the  laser fluence value is crucial not only to induce the solid-liquid transition required for convective processes, but also to control the final surface topography and nanostructure shapes. Thus, laser fluence is a second key factor in controlling surface organization on the nanoscale.

The laser dose is the third parameter exploited to guide the final patterns. Fig.\ref{fgrexp}(C) reveals the essential role of $N_{DPS}$ in nanostructures formation and transient organization. Thus, the laser fluence was fixed here at 0.18 J/cm$^{2}$ and $\Delta t=8$ ps. Results obtained with $N_{DPS}$ varying between 0 and 30 pulses are presented to indicate its central role in nanostructures self-organization. For 1 $\leq N_{DPS}\leq 10$, there is no nanostructure observed and higher number of double-pulses sequences are required, as the total irradiated fluence was below surface modification or nanostructures formation threshold. However, for $10$ $\leq N_{DPS}\leq 15$, some nanoroughness have appeared on the irradiated surface.  By increasing the number of double-pulses sequences to $20$ $\leq N_{DPS}\leq 25$, we started to observe emergence of organized small peaks, and their concentration and height increase progressively at  $N_{DPS} = 25$. Furthermore, while increasing the number of double-pulses sequences to 30 $\leq N_{DPS}$, the peaks concentration has decreased and other structures developed.

\section{Appendix B: The Swift-Hohenberg equation}
\paragraph{Swift-Hohenberg equation as a model of Rayleigh-Bénard convection}
The Swift Hohenberg equation was introduced in~\cite{swift1977hydrodynamic} as a model of Rayleigh-Bénard convection. It can be written adimensionally as \(\dot{u} = \epsilon \tilde{u} -(1+\nabla^{2})^{2} \tilde{u} + \mathcal{N}[\tilde{u}]\), where \(\tilde{u}(x,y,t)\) is a real scalar field, \(\mathcal{N}[\tilde{u}]\) is a nonlinear term, and \(\epsilon>0\) is a \emph{bifurcation parameter}. Commonly, \(\mathcal{N}[\tilde{u}]=\tilde{u}^3\) which implies that perturbations of the \(\tilde{u}_{b}=0\) solution are selectively amplified, as only the Fourier modes whose wave vector norm lies on a certain interval centered at on \(1\) with width depending on \(\epsilon\), will be amplified, all others being attenuated (see fig.\ref{fgr2}). The left endpoint of this interval not being zero, it is called a type-I-s instability~\cite{cross1993pattern}. Selective amplification of perturbations depending on the wave number leads to the formation of \emph{patterns}, which, since the selection of wave vectors depends only on the norm, show no preferential direction.

\begin{figure}[h!]
\includegraphics[width=8.5 cm]{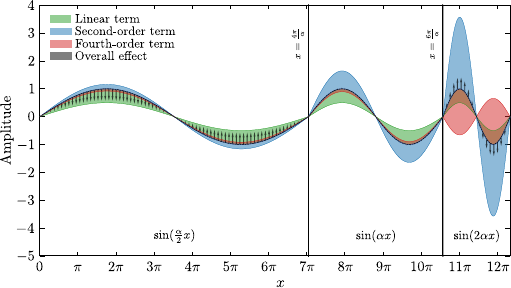}
\caption{\label{fgr2} Action of the linear part of the one-dimensional Swift Hohenberg equation $\partial_t \tilde{u} = \epsilon \tilde{u} -(1+\partial_x^2)^2\tilde{u} + \mathcal{N}[\tilde{u}]$ on three Fourier modes $u = a(t)\sin(kx)$. For small $a(t)$, ignoring the nonlinear part and to first order in time, $\partial_t \tilde{u} = (\epsilon -1)+2k^2-k^4)a(t)\sin(kx)$. 
The zeros of the factor on the right hand-side are $k=\sqrt{1\pm\sqrt{1+\left(\epsilon-1\right)}}$. For $\epsilon=0.5$ and $k_1 =\sqrt{1-\sqrt{0.5}}$ the leftmost zero, the overall derivative of the mode $\sin(k_1 x)$ is zero, as can be seen on the center plot. Modes with lower frequency will be attenuated, as can be seen on the leftmost plot, where we show the magnitude of the various terms on $\sin(0.5 k_1x)$. Finally, the mode $\sin(2 k_1x)$, for which the frequency lies between the zeros of the factor is amplified, as can be seen on the rightmost plot.}
\end{figure}
\paragraph{SH can be derived on symmetry grounds alone}
Given appropriate boundary conditions (e.g. periodic) SH is the simplest equation with a type-I-s instability that is isotropic, invariant with respect to translations and to field inversion \(\tilde{u}\rightarrow -\tilde{u}\)~\cite{cross1993pattern}. For a system such as ours, that satisfies these symmetries, it can thus be used as a model of pattern formation, regardless of specificities. We do break the symmetry with respect to \(\tilde{u}\rightarrow -\tilde{u}\), by setting \(\mathcal{N}[\tilde{u}] = \gamma \tilde{u}^2 - \tilde{u}^{3}\), obtaining:
\begin{align}
    \label{eq:shadim}
    \dot{\tilde{u}} = \epsilon \tilde{u} - (1+ \tilde{\nabla}^2)^2\tilde{u} + \gamma \tilde{u}^2 - \tilde{u}^3,
\end{align}
the quadratic term allowing small amplitude destabilization and the existence of the hexagonal patterns which we observe experimentally, with the negative cubic term, which dominates for large amplitudes, still controlling the magnitude of the instabilities, which would otherwise grow without bound.

\paragraph{Rescaling to obtain the adimensional version}
Consider the SH equation introduced in~\cite{swift1977hydrodynamic}, modified to include a $u^2$ term:

\begin{align}
\label{eq:shdim}
    \tau_0 \dot{u} = \epsilon_0 u - \xi_0^4 \left( q_0^2 + \nabla^2 \right)^2u+ \gamma_0 u^{2}-g_0u^3
\end{align}

where \(u(x,y,t)\) is a real scalar field, \(\tau_0\) is the \emph{characteristic time scale} of the instability and has dimensions of time, \(\epsilon_0\) is a dimensionless \emph{control parameter} chosen such that \(\epsilon_0=0\) defines the onset of instability, \(\xi_0\) is the \emph{coherence length} determining the spacial range over which local disruptions (boundary or topological defects) perturb the surrounding pattern, \(q_0\) is the \emph{critical wave number} (the wave number that gets selected at onset), \(\gamma_0\) is a parameter with dimensions of \(u^{-1}\) determining the strength of the \(u^2\) nonlinearity, which is the only term of the equation which is not odd and thus breaks invariance with respect to the transformation \(u\rightarrow -u\), and \(g_0\) controls the magnitude of the cubic terms and has dimensions of \(u^{-2}\). Throughout this letter we assume periodic boundary conditions. It is clear by inspection that equation \ref{eq:shadim} can be obtained from \ref{eq:shdim} upon a rescaling of time $\tilde{t} = a^4 \tau_0 t$, length \footnote{Assuming for simplicity that the length scales in the \(x\) and \(y\) directions are the same.} $\tilde{x} = \pm\frac{\lambda_0}{2\pi}x ,\, \tilde{y}=\pm\frac{\lambda_0}{2\pi}y$, and magnitude $\tilde{u} = \pm \sqrt{g_0} a^2 u$, with $\epsilon =a^4 \epsilon_0$ and $\gamma = \sqrt{g_0}a^6\gamma_0$ and $a=\frac{\lambda_0}{2 \pi\xi_0}$.

\begin{figure}[h!]
\includegraphics[width=8.5 cm]{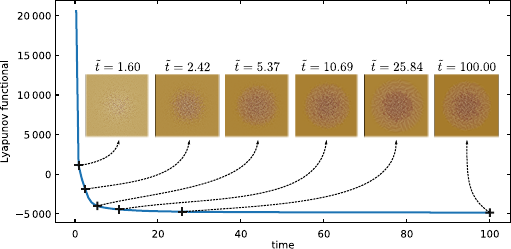}
\caption{\label{fig:laypvstime} Illustration of the time evolution of the Lyapunov functional \(\mathcal{L} \left[ \tilde{u} \right] = \int_{\Omega} \frac{\tilde{u}}{2}\left( \nabla^4\tilde{u} + 2\nabla^2\tilde{u}+\tilde{u} \right)+\frac{1}{4}\tilde{u}^4 -\frac{\gamma}{3} \tilde{u}^3-\frac{\epsilon}{2}\tilde{u}^2 d\mathbf{x}\). Inset images represent the field at given solver times. $\epsilon$ is a 2D Gaussian ramp centered at the origin of the image, to mimic the laser fluence distribution.}
\end{figure}

\paragraph{SH has rich pattern solutions but simple dynamics}
In spite of the complexity of its pattern solutions, SH has a simple dynamics (with the periodic boundary conditions on the domain \(\Omega\) used throughout the paper): indeed, a Lyapunov functional \(\mathcal{L}[\tilde{u}]\) exists such that \(\dot{\tilde{u}} = - \frac{\delta V}{\delta \tilde{u}}\). For \(\mathcal{N}[\tilde{u}] = \gamma \tilde{u}^2 - \tilde{u}^{3}\), using the Euler-Lagrange equations we can show that \(\mathcal{L} \left[ \tilde{u} \right] = \int_{\Omega} \frac{\tilde{u}}{2}\left( \nabla^4\tilde{u} + 2\nabla^2\tilde{u}+\tilde{u} \right)+\frac{1}{4}\tilde{u}^4 -\frac{\gamma}{3} \tilde{u}^3-\frac{\epsilon}{2}\tilde{u}^2 d\mathbf{x}\). This potential decreases during the dynamics and, since it is bounded below, it converges asymptotically to a stable value~\cite{cross1993pattern} as shown in Fig.\ref{fig:laypvstime}.

\section{Appendix C: Machine Learning model\label{sec:ML-model}} 

Consider a physical field \(u(x,t)\) which is mainly the result of a certain physical process, described in terms of unknown initial conditions $u_{0}$ and physical process parameters $\varphi$, using a PDE \(\dot{u}=f(\varphi,u_0)\)\footnote{Throughout this section we assume that $u$ is a real field and periodic boundary conditions; we further assume that conditions are satisfied such that $u$ is unique (existence is posited since we are assuming that the process is modeled by this equation and we observe the fields)}. Assume that a certain \emph{observed} quantity \(\theta\) affects $\varphi$ and possibly other unknown latent variables, which may in turn affect $u$, and experimental data \({\left\{ \theta^i,u^i \right\}}_{i=1\ldots N }\), where $N$ is small. We would like to sample from the distribution of \(u\) given \(\theta\). Unfortunately, since $u_0$ is unknown, $p(u|\theta)$ cannot be calculated directly:
\begin{align*}
p(u|\theta) &= \sum_{u_0}\sum_{\varphi}p(u|u_0, \theta,\varphi)p(u_0,\varphi|\theta)
\end{align*}
The quantity of physical interest is often not \(u\), however, but a certain function \(F(u)\), where \(F(u)\) is typically simpler than \(u\). With this in mind, assume that $F$ is such that $p(F(u)|u_0, \theta, \varphi)=p(F(u)|\theta,\varphi)$, and further assume $p(\varphi|\theta)$ is peaky, in the sense that there is a $\tilde{\varphi}$ such that $p(\varphi|\theta)=0$ for $\varphi\neq \tilde{\varphi}$ (if $p$ is continuous, in particular, $p(\varphi|\theta) = \delta(\varphi - \tilde{\varphi}(\theta))$. The first assumption implies $p(F(u)|\theta) = \sum_{u_0}\sum_{\varphi}p(F(u)| \theta,\varphi)p(u_0,\varphi|\theta) = \sum_{\varphi}p(F(u)| \theta,\varphi)p(\varphi|\theta)$. The second assumption implies that \(\tilde{\varphi}\) is a function of \(\theta\), which implies that \(\theta\) is at least locally a function of \(\tilde{\varphi}\) and we can write $p(F(u)|\theta) = p(F(u)| \tilde{\varphi})$. 

To learn the relationship between \(F(u)\) and \(\tilde{\varphi}\) from experimental data we can parameterize \(\tilde{\varphi}\) with a Neural Network (\(\alpha\)), for example, and maximize the log likelihood of observing the \(F(u^{i})\):
\begin{align}
\label{eq:learn-via-max-log-likelihood-obs}
\bar{\alpha}=\argmax \sum\limits_{i=1}^{N}\log p \left( F(u^{i})|\tilde{\varphi}(\theta^{i});\alpha \right)
\end{align}
If \(F(u)\) is high-dimensional, this relationship is potentially complex, requiring \(N\) large to model satisfactorily. Having physical knowledge in the form of a differential equation solver \(u=\mathrm{Solver}(\tilde{\varphi},u_0)\), considerably simplifies the problem, since in feature space, for a process satisfying the conditions above, we can choose arbitrary $u_0$ and fit only the relationship between \(\theta\) and the \emph{parameters} of the PDE, which generally have much lower dimension than the field \(F(u)\). This being a simpler task, we expect that a smaller \(N\) will suffice to produce a satisfactory model. Assuming data is generated i.i.d. from a fixed-variance Gaussian distribution, this objective corresponds to minimizing the mean squared error between experimental images and generated images, in feature space
\begin{align*}
\bar{\alpha}=\argmin \frac{1}{N}\sum\limits_{i=1}^{N} \left\|F(u^{i})-F(\mathrm{Solver}(\tilde{\varphi}_{\alpha}(\theta^{i}), u_{0}))  \right\|^{2}
\end{align*}
for arbitrary \(u_{0}\). The problem can be further simplified by noting that $\log p(F(u)| \theta(\tilde{\varphi}),\tilde{\varphi})p(\tilde{\varphi}|\theta) = \log p(F(u)|\tilde{\varphi})+\log p(\tilde{\varphi}|\theta)$. With \(\tilde{\varphi}_{1}\) the maximizer of the first term, \(\tilde{\varphi}_{1}= \argmax_{\tilde{\varphi}} \log p(F(u)|\tilde{\varphi})\), then $\max_{\tilde{\varphi}}\left\{\log p(F(u)|\tilde{\varphi})+\log p(\tilde{\varphi}|\theta)\right\}\geq  \log p(F(u)|\tilde{\varphi}_{1})+\log p(\tilde{\varphi}_{1}|\theta)$.
Provided we find \(\tilde{\varphi}_1^{i}\) for each \(u^{i}\), we can replace the log likelihood maximization objective with that of maximizing a lower bound $\bar{\alpha} = \argmax_{\alpha}\sum\limits_{i=1}^N \log p \left( \tilde{\varphi}^{i}|\theta^{i};\alpha \right)$ which, repeating the argument above, corresponds to minimizing the mean squared error
\begin{align*}
    \bar{\alpha}=\argmin_{\alpha}\frac{1}{N}\sum\limits_{i=1}^N \left\|\tilde{\varphi}^i-\tilde{\varphi}_{\alpha}(\theta^{i})  \right\|^{2}
\end{align*}

Having an efficient solver and assuming sufficient regularity, one can pre-generate a number of fields \(\mathcal{U}_p = {\left\{ u_{p}^{k} \right\}}_{k=1\ldots M}\) such that the expected distance to the nearest neighbor in feature space is as small as one would like. Assuming that we can pre-generate fields at will, \(\delta\rightarrow 0\), which implies that we can we can set \(\tilde{\varphi}^{i}\) as the solver parameter of the nearest neighbor in feature space, among pre-generated fields, of \(u^{i}\). By integrating physical information in this way, we are able to learn the desired relationship using any method suited to few data, such as support vector regression.
\paragraph{Choosing the feature transformation $F$}
The feature mapping $F$ is clearly not unique, as if $F$ is a suitable feature mapping so is $aF +b$ for real $a,b$, for example. Given a sufficiently detailed model, it could be derived  from first principles or, given enough data and for simple dynamics, it could be learned~\cite{greydanus19_hamil_neural_networ}. In a partial physical model/few data setting, however, we are forced to \emph{choose} a transformation that will have the desired properties of simplification and separation. In a context of optimization where we would have a specific pattern feature of interest, traditional image features could be an appropriate choice, but in our pattern discovery setting they could limit discriminating power for unknown patterns. To keep $F$ as unconstrained as possible, we choose it to be learned automatically from data for a complex classification task on a large and diverse dataset. Specifically, we choose $F$ as a deep convolutional neural network~\cite{lecun1998gradient} (CNN) pretrained for a broad classification task on Imagenet~\cite{deng09_imagen}, which contains 14 million examples. CNNs are translation equivariant, which makes them suited for a pattern specification task, and learned features retain scale information~\cite{graziani21_scale_invar_state_art_cnns_train_imagen} provided we section them before the last layer. The quality of the feature transformation can be evaluated by combining cross-validation results (to do so, we trained a neural network surrogate of the solver in feature space) and the expert-evaluated accuracy of a kNN clustering task.  

\section{Appendix D: Pseudospectral solver}
\paragraph{Strang-splitting pseudospectral solver}
The ML workflow requires the unsupervised generation of many accurate numerical solutions of the SH equation. To that effect, we implemented a second-order split-step~\cite{leveque2007finite,yoshida1990construction} pseudo-spectral solver providing a good compromise between accuracy and speed. In the sequel, we write SH as \(\dot{u} = \mathbf{L}[u] + \mathbf{N}[u]\) with linear \(\mathbf{L}\) and a non-linear \(\mathbf{N}\) differential operators acting on \(u\) for clarity. We use \textit{Strang splitting}~\cite{strang1968construction} \footnote{We examined a symmetric 3-fold Strang composition method~\cite{yoshida1990construction} of order four, but the improvement was not sufficient for our purposes to justify the increase in execution time, which stems not only from the extra time stepping, but from the higher-order methods for each of the individual steps.} a split-step second order time stepping method, with order two methods for the individual steps: Ralston's for the nonlinear part; the Trapezoidal rule for the linear part in Fourier space, computed using FFT~\cite{cooley1965algorithm}, allowing us to take this step at quasi-linear time complexity. With \(T(\mathbf{L}, u^t,k)\) and \(R(\mathbf{N}, u^t,k)\) representing the Trapezoidal and Ralston's numerical methods solving, respectively, \(\dot{u} = \mathbf{L} [u]\) and \(\dot{u} = \mathbf{N} [u]\) over a time step of \(k\) starting with data \(u^t\), \(\mathcal{F}\) denoting the discrete Fourier transform, and \(\mathcal{F}[u]=v\), our scheme reads \(u^* = \mathcal{F}^{-1} \left\{ T(\mathbf{L},v^t, \frac{dt}{2}) \right\},\quad u^{**} = R(\mathbf{N}, u^*, dt),\quad u^{t+1} =  \mathcal{F}^{-1} \left\{ T(\mathbf{L}, v^{**}, \frac{dt}{2})\right\}\)
where \(u^*, u^{**}\) are intermediate solutions. Note that that periodic boundary conditions are automatically enforced by the Fourier transform.
\paragraph{Stability control}
Stability of the split step method is determined by that of the explicit step, since the region of absolute stability of the Trapezoidal rule includes the entire left half of the complex plane, making it suitable for the solution of stiff equations~\cite{butcher2016numerical,leveque2007finite}. We use an adaptive time step of the order of inverse of the spectral norm of the Jacobian of the nonlinear stepping operator

and control for divergence by checking the derivative of the Lyapunov functional at fixed iteration intervals. 

\section{Appendix F: Correlation between SH parameter predictions}
\begin{figure}[h!]
\includegraphics[width=8.6 cm]{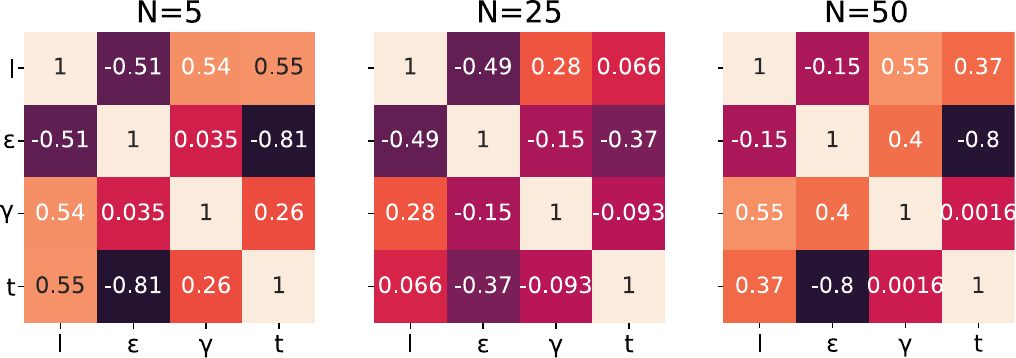}
\caption{Pearson correlation between predicted SH parameters $l, \epsilon,\gamma, t$ for N=5,25,50. }
\label{fig:sh-param-corr}
\end{figure}

The predictions of the different SH parameters $\varphi = l,\epsilon, \gamma, \tilde{t}$ by the ML model trained on the experimental data according to the method described in this letter are correlated for each value of $N$. To see that, for each value of $N$ and for predicted SH parameter, we calculate the Pearson correlation coefficient between pairs of the 501² point prediction plots in Fig. 4 in the letter: covariance divided by the product of standard deviations. For ease of comparison, we standardize each image by subtracting the mean and scaling to unit variance. In this case, the Pearson coefficient is a measure of the sample covariance between prediction images. For SH parameters $X,Y$:
\begin{align}
    \rho(X,Y) = \frac{1}{501^2-1}\sum_{i=1}^{501^2} (x_i-\bar{x})(y_i-\bar{y}).
\end{align}We present a heatmap with calculated correlations for all parameter pairs for $N=5,25,50$ in Fig.~\ref{fig:sh-param-corr}.

\section{\label{sec:time-scale}Appendix G: Timescale of the convective instability}
We now show that the timescale of convective instability $\tau_0$ is consistent with the transversal dissipation time of  \(\approx 10\) ps reported in~\cite{Nakhoul2021Apr} by combining experimental information with that obtained via our ML model. Our experimental design does not provide direct access to $\tau_{0}$, which is typically found by examining the growth rate of the patterns close to onset. For small \(u,\epsilon_0\), and for \(q\approx q_{0}\), the growth rate of the linear instability of wave number \(q\) of SH is \(\sigma(q) \approx \frac{1}{\tau_0} \left[ \epsilon_0 - \xi_0^2(q-q_0)^2 \right]\), implying \(\tau_0^{-1}= \left.\frac{\partial \sigma}{\partial \epsilon_0} \right|_{\epsilon_0=0, q=q_0}\). For a perturbation of $u=0$, for small \(\Delta t\) and close to onset, \(\Delta u\) is mostly due to the growth of the $q_0$ mode. Hence, to first order, $\frac{\Delta u}{\Delta t} = \sigma(q_0,\epsilon_0) u_{q_0}(0)$. Assuming \(u_{q_0}(0)\) is of the same order of magnitude throughout the domain and small compared to \(u_{q_0}(\epsilon_0, \Delta t)\), we have \(\sigma(q_0,\epsilon_0) \approx \frac{u_{q_0}(\epsilon_0,\Delta t)}{ u_{q_0}(0)\times\Delta t}\). Since $\tau_0^{-1}\approx \frac{\Delta \sigma}{\Delta \epsilon_{0}}$ we obtain $\tau_0= \frac{u}{\Delta u} \Delta t\Delta\epsilon\, a^4$. $\Delta t$ is of the order of $10$ ps, and we are certain that some evolution of $u$ already took place. For $\tau_{0}$ to be of the same order of magnitude $\frac{u}{\Delta u}\Delta\epsilon\, a^4$ must be order one. To remain in the limit of small $u$ where this approximation is valid, \(u/\Delta u\) is of order one. $a^4\sim 1-10^{4}$, assuming $\eta_0$ small compared to $\lambda_0$, which implies a change in \(\epsilon\) of the order of \(10^{-4}\) and \(10^{0}\). We recover the small \(\epsilon\) limit, which we did \emph{not} impose specifically, which is evidence of agreement between SH and experimental results. Moreover, the simulation time for our solver for experimental patterns being of order $1$ to \(10^{3}\), assuming this corresponds to the resolidification time of the fluid material \(\approx 100\) ps~\cite{Nakhoul2021Apr}, we obtain a characteristic time in the range of $10^{-10}$ to $10^{-17}$ s, which is consistent with the transversal dissipation time of \(\approx 10\) ps reported in~\cite{Nakhoul2021Apr}.
\end{appendices}
\end{document}